\documentclass[a4paper]{jpconf}
\usepackage{graphicx}
\usepackage{amsmath,amssymb,amsfonts,amsthm}\newcommand{\be}{\begin{equation}}
\newcommand{\ee}{\end{equation}}
\newcommand{\beq}{\begin{eqnarray}}
\newcommand{\eeq}{\end{eqnarray}}

\begin{document}
\title{Cosmological consequences of the noncommutative spectral geometry
  as an approach to unification}
                
\author{Mairi Sakellariadou}

\address{Department of Physics, King's College, University of
  London, Strand WC2R 2LS, London, U.K.}

\ead{mairi.sakellariadou@kcl.ac.uk}

\begin{abstract}
Noncommutative spectral geometry succeeds in explaining the physics of
the Standard Model of electroweak and strong interactions in all its
details as determined by experimental data. Moreover, by construction
the theory lives at very high energy scales, offering a natural
framework to address early universe cosmological issues. After
introducing the main elements of noncommutative spectral geometry, I
will summarise some of its cosmological consequences and discuss
constraints on the gravitational sector of the theory.

\end{abstract}

%%%%%%%%%%%%%
\section{Introduction}
Approaching Planckian energies, gravity can no longer be considered as
a classical theory, the quantum nature of space-time becomes apparent
and geometry can no longer be described in terms of Riemannian
geometry and General Relativity.  At such energy scales, all forces
should be unified so that all interactions correspond to one
underlying symmetry. The nature of space-time would change in such a
way, so that its low energy limit implies the diffeomorphism and
internal gauge symmetries, which govern General Relativity and gauge
groups on which the Standard Model is based, respectively.  Thus, near
Planckian energies one should search for a formulation of geometry
within the quantum framework.  Such an attempt has been realised
within NonCommutative Geometry (NCG)~\cite{ncg-book1, ncg-book2}.
Even though this approach is still in terms of an effective theory, it
has already led to very encouraging results and is now at a stage to
be confronted with experimental and observational data.

Since all physical data are of a spectral nature, we will use the
notion of a spectral triple, which is analogous to Fourier transform
we are familiar with in commutative spaces. Distances are measured in
units of wavelength of atomic spectra, while the notion of a real
variable, taken as a function on a set $X$, will be replaced by a
self-adjoint operator $D$ in a Hilbert space.  The space $X$ will be
described by the algebra ${\cal A}$ of coordinates, represented as
operators in a fixed Hilbert space ${\cal H}$.  Thus, the usual
emphasis on the points of a geometric space is replaced by the
spectrum of an operator and the geometry of a noncommutative space is
determined in terms of the spectral triple $({\cal A},{\cal H},D)$.

Noncommutative geometry spectral action, in its current version which
is still at the classical level and it certainly considers almost
commutative spaces, it nevertheless offers~\cite{ccm} an elegant
geometric interpretation of the most successful phenomenological model
of particle physics, namely the Standard Model (SM). Certainly if the
Large Hadron Collider (LHC) supports evidence for physics beyond the
SM, one should consider extensions of the effective theory we will
adopt here.  Closer to the Planck era one should consider the full,
still unknown, theory, while at even higher energy scales the whole
concept of geometry may lose its familiar meaning. Nevertheless, the
effective theory we will consider here lives, by construction, at very
high energies, offering an excellent framework to address open
questions of early universe cosmology~\cite{m2010}.

In what follows, I will briefly review the main elements of the NCG
spectral action as an approach to unification and then discuss some of
its cosmological consequences~\cite{ m2010,Nelson:2008uy,
  Nelson:2009wr, mmm, wjm1, wjm2}.

%%%%%%%%%%%%%%%%%%%%%%%
\section{NCG spectral action}
In the context of noncommutative geometry, all information about a
physical system is contained within the algebra of functions,
represented as operators in a Hilbert space, while the action and
metric properties are encoded in a generalised Dirac operator.  The
geometry is specified by a spectral triple $({\cal A},{\cal H},D)$,
defined by an algebra ${\cal A}$, a Hilbert space ${\cal H}$ and a
generalised Dirac operator $D$.

Adopting the simplest generalisation beyond commutative spaces, we
take the extension of our smooth four-dimensional manifold, ${\cal
  M}$, by taking its product with a discrete noncommuting manifold
${\cal F}$.  Thus, ${\cal M}$ describes the geometry for the
four-dimensional space-time, while the noncommutative space ${\cal F}$
specifies the internal geometry for the SM and is composed of just two
points.  Considering the Standard Model of electroweak and strong
interactions as a phenomenological model, we will look for a geometry,
such that the associated action functional leads to the SM with all
specifications as determined by experimental data.

Within NCG, a space described by the algebra of real coordinates is
represented by self-adjoint operators on a Hilbert space.  Since real
coordinates are represented by self-adjoint operators, all information
about a space is encoded in the algebra of coordinates ${\cal A}$,
which is the main input of the theory. Under the assumption that the
algebra constructed in ${\cal M}\times {\cal F}$ is
symplectic-unitary, it turns out that ${\cal A}$ must be of the form
\begin{equation}
\mathcal{A}=M_{a}(\mathbb{H})\oplus M_{k}(\mathbb{C})~,
\end{equation}
with $k=2a$ and $\mathbb{H}$ denoting the algebra of quaternions.  The
choice $k=4$ is the first value that produces the correct number of
fermions, namely $k^2=16$, in each of the three
generations~\cite{Chamseddine:2007ia}. The number of generations is a
physical input. Certainly if at LHC new particles are discovered, one
may be able to accommodate them by including a higher value for the
even number $k$. 

The operator $D$ corresponds to the inverse of the Euclidean
propagator of fermions, and is given by the Yukawa coupling matrix
which encodes the masses of the elementary fermions and the
Kobayashi--Maskawa mixing parameters.  The fermions of the SM provide
the Hilbert space ${\cal H}$ of a spectral triple for the algebra
${\cal A}$, while the bosons of the SM are obtained through inner
fluctuations of the Dirac operator of the product ${\cal M}\times
{\cal F}$ geometry.

The spectral action principle states that the bosonic part of the
spectral functional $S$ depends only on the spectrum of the Dirac
operator and its asymptotic expression, and for large energy $\Lambda$
is of the form ${\rm Tr}(f(D/\Lambda))$, with $f$ being a cut-off
function, whose choice plays only a small r\^ole.  The physical
Lagrangian has also a fermionic part, which has the simple linear form
$(1/2)\langle J\psi, D\psi\rangle$, where $J$ is the real structure on
the spectral triple and $\psi$ are spinors defined on the Hilbert
space~\cite{sp-act}.  Applying the spectral action principle to the
inner fluctuations of the product geometry ${\cal M}\times {\cal F}$,
one recovers the Standard Model action coupled to Einstein and Weyl
gravity plus higher order nonrenormalisable interactions suppressed by
powers of the inverse of the mass scale of the theory~\cite{ccm}.

To study the implications of this noncommutative approach coupled to
gravity for the cosmological models of the early universe, one can
concentrate just on the bosonic part of the action; the fermionic part
is however crucial for the particle physics phenomenology of the
model. 

Using the heat kernel method, the bosonic part of the spectral action
can be expanded in powers of the scale $\Lambda$ in the
form~\cite{ccm,ac1996,ac1997}
\begin{equation}
\label{eq:sp-act}
{\rm Tr}\left(f\left(\frac{D}{\Lambda}\right)\right)\sim 
\sum_{k\in {\rm DimSp}} f_{k} 
\Lambda^k{\int\!\!\!\!\!\!-} |D|^{-k} + f(0) \zeta_D(0)+ {\cal O}(1)~,
\end{equation}
with the momenta $f_k$ of the cut-off function $f$ given by
\be\label{eq:moments0}
f_k\equiv\int_0^\infty f(u) u^{k-1}{\rm d}u\ \ \mbox{for}\ \ k>0 ~~,
\ \ \mbox{and}\ \ f_0\equiv f(0)~;
\ee
 the noncommutative integration is defined in terms of residues of
 zeta functions, $\zeta_D (s) = {\rm Tr}(|D|^{-s})$ at poles of the
 zeta function, and the sum is over points in the {\sl dimension
   spectrum} of the spectral triple.

In this way, one obtains a Lagrangian which contains in addition to
the full SM Lagrangian, the Einstein-Hilbert action with a
cosmological term, a topological term related to the Euler
characteristic of the space-time manifold, a conformal Weyl term and a
conformal coupling of the Higgs field to gravity.  Writing the
asymptotic expansion of the spectral action, a number of geometric
parameters appear, which describe the possible choices of Dirac
operators on the finite noncommutative space. These parameters
correspond to the Yukawa parameters of the particle physics model and
the Majorana terms for the right-handed neutrinos.  The Yukawa
parameters run with the Renormalisation Group Equations (RGE) of the
particle physics model. Since running towards lower energies, implies
that nonperturbative effects in the spectral action cannot be any
longer neglected, any results based on the asymptotic expansion and on
renormalisation group analysis can only hold for early universe
cosmology. For later times, one should instead consider the full
spectral action.

More precisely, the bosonic action in Euclidean signature
reads~\cite{ccm}
\beq\label{eq:action1} 
{\cal S}^{\rm E} = \int \left(
\frac{1}{2\kappa_0^2} R + \alpha_0
C_{\mu\nu\rho\sigma}C^{\mu\nu\rho\sigma} + \gamma_0 +\tau_0 R^\star
R^\star
\right.  
%\nonumber\\
+ \frac{1}{4}G^i_{\mu\nu}G^{\mu\nu
  i}+\frac{1}{4}F^\alpha_{\mu\nu}F^{\mu\nu\alpha}\nonumber\\ 
+\frac{1}{4}B^{\mu\nu}B_{\mu\nu}
%\nonumber\\ 
+\frac{1}{2}|D_\mu{\bf H}|^2-\mu_0^2|{\bf H}|^2
%\nonumber\\ 
\left.
- \xi_0 R|{\bf H}|^2 +\lambda_0|{\bf H}|^4
\right) \sqrt{g} \ d^4 x~, \eeq
where 
\beq\label{bc} 
\kappa_0^2&=&\frac{12\pi^2}{96f_2\Lambda^2-f_0\mathfrak{c}}
~,\nonumber\\
\alpha_0&=&-\frac{3f_0}{10\pi^2}~,\nonumber\\ 
\gamma_0&=&\frac{1}{\pi^2}\left(48f_4\Lambda^4-f_2\Lambda^2\mathfrak{c}
+\frac{f_0}{4}\mathfrak{d}\right)~,\nonumber\\ 
\tau_0&=&\frac{11f_0}{60\pi^2}~,\nonumber\\
\mu_0^2&=&2\Lambda^2\frac{f_2}{f_0}-{\frac{\mathfrak{e}}{\mathfrak{a}}}~,
\nonumber\\
\xi_0&=&\frac{1}{12}~,\nonumber\\
\lambda_0&=&\frac{\pi^2\mathfrak{b}}{2f_0\mathfrak{a}^2}~,
\eeq
with $\mathfrak{a}, \mathfrak{b}, \mathfrak{c}, \mathfrak{d},
\mathfrak{e}$ given by~\cite{ccm}
\beq\label{eq:Ys-oth}
 \mathfrak{a}&=&{\rm Tr}
\left( Y^\star_{\left(\uparrow 1\right)} Y_{\left(\uparrow 1\right)} +
Y^\star_{\left(\downarrow 1\right)} Y_{\left(\downarrow
  1\right)}
%\right.  \nonumber\\ &&
+ 
%\left. 
3\left( Y^\star_{\left(\uparrow 3\right)} Y_{\left(\uparrow 3\right)}
+ Y^\star_{\left(\downarrow 3\right)} Y_{\left(\downarrow 3\right)}
\right)\right)~,\nonumber\\ \mathfrak{b}&=&{\rm Tr}\left(\left(
Y^\star_{\left(\uparrow 1\right)} Y_{\left(\uparrow 1\right)}\right)^2
+ \left(Y^\star_{\left(\downarrow 1\right)} Y_{\left(\downarrow
  1\right)}\right)^2
%\right.
%\nonumber\\ 
%&&
+ 
%\left. 
3\left( Y^\star_{\left(\uparrow 3\right)} Y_{\left(\uparrow
  3\right)}\right)^2 + 3 \left(Y^\star_{\left(\downarrow 3\right)}
Y_{\left(\downarrow 3\right)}
\right)^2\right)~,\nonumber\\ \mathfrak{c}&=&{\rm Tr}\left(Y^\star_R
Y_R\right)~,\nonumber\\ \mathfrak{d}&=&{\rm Tr}\left(\left(Y^\star_R
Y_R\right)^2\right),\nonumber\\ \mathfrak{e}&=&{\rm Tr}\left(Y^\star_R
Y_RY^\star_{\left(\uparrow 1\right)} Y_{\left(\uparrow
  1\right)}\right)~, \eeq
with $Y_{\left(\downarrow 1\right)}, Y_{\left(\uparrow 1\right)},
Y_{\left(\downarrow 3\right)}, Y_{\left(\uparrow 3\right)}$ and $Y_R$
being $(3\times 3)$ matrices, with $Y_R$ symmetric; the $Y$ matrices
are used to classify the action of the Dirac operator and give the
fermion and lepton masses, as well as lepton mixing, in the asymptotic
version of the spectral action.  Note that ${\bf H}$ is a rescaling of
the Higgs field, so that the kinetic terms are normalised.  One should
be cautious that the relations in Eq.~(\ref{bc}) are only valid at
unification scale $\Lambda$; it is incorrect to consider them as
functions of the energy scale.

The noncommutative spectral geometry approach leads to various
phenomenological consequences.  Normalisation of the kinetic terms
implies
\be
{g_3^2f_0\over 2\pi^2}={1\over 4} ~~\mbox{and}~~ g_3^2=g_2^2={5\over
  3}g_1^2\nonumber~, \ee
while
\be \sin^2\theta_{\rm W}=\frac{3}{8}~; \ee
a relation which holds also for SU(5) and SO(10), while assuming the
{\sl big desert} hypothesis, one can find the running of the three
couplings $\alpha_i=g_i^2/(4\pi)$.  One-loop RGE for the running of
the gauge couplings and the Newton constant, shows that they do not
meet exactly at one point, the error is though within just few
percent.  Therefore, the model in its simplified form, does not
specify a unique unification energy, it however leads to the correct
representations of the fermions with respect to the gauge group of the
SM, the Higgs doublet appears as part of the inner fluctuations of the
metric, and Spontaneous Symmetry Breaking mechanism arises naturally
with the negative mass term without any tuning. Moreover, the see-saw
mechanism is obtained, the 16 fundamental fermions are recovered, and
a top quark mass of $M_{\rm top}\sim 179 ~{\rm Gev}$ is predicted.
The mass of Higgs in zeroth order approximation of the spectral action
is $\sim 170 {\rm GeV}$, which is strictly speaking ruled out by
current experimental data. Nevertheless, the result depends on the
value of gauge couplings at unification scale, which is certainly
uncertain, while it was found neglecting the nonminimal coupling
between the Higgs field and the Ricci curvature. It is however worth
noticing that the NCG approach leads to the correct order of magnitude
for the Higgs mass, a result which was not obvious {\sl a priori}.

%%%%%%%%%%%%%%%%%%%%%%%%%%%%
\section{Cosmological consequences}
To use the formalism of spectral triples in NCG, it is convenient to
work with Euclidean rather than Lorentzian signature.  Thus, the
analysis of the cosmological consequences of the theory relies on a
Wick rotation to imaginary time, into the Lorentzian signature.

The Lorentzian version of the gravitational part of the asymptotic
formula for the bosonic sector of the NCG spectral action, including
the coupling between the Higgs field and the Ricci curvature scalar,
reads~\cite{ccm}
\be\label{eq:1.5} {\cal S}_{\rm grav}^{\rm L} = \int \left(
%\frac{1}{16\pi G} R 
\frac{1}{2\kappa_0^2} R + \alpha_0
C_{\mu\nu\rho\sigma}C^{\mu\nu\rho\sigma} + \tau_0 R^\star
R^\star
%\right.  \nonumber\\ -\left.  
\xi_0 R|{\bf H}|^2 \right)
\sqrt{-g} \ d^4 x~, \ee
leading to the equations of motion~\cite{Nelson:2008uy}
\be\label{eq:EoM2} R^{\mu\nu} - \frac{1}{2}g^{\mu\nu} R +
\frac{1}{\beta^2} \delta_{\rm cc}\left[
  2C^{\mu\lambda\nu\kappa}_{;\lambda ; \kappa} +
  C^{\mu\lambda\nu\kappa}R_{\lambda \kappa}\right]
\nonumber\\ = 
%\ 8\pi G
\kappa_0^2 \delta_{\rm cc}T^{\mu\nu}_{\rm matter}~, \ee
with
\be
\beta^2 \equiv -\frac{1}{4\kappa_0^2 \alpha_0}
 \ \ \mbox{and}\ \ 
\delta_{\rm cc}\equiv[1-2\kappa_0^2\xi_0{\bf H}^2]^{-1}~.
\ee
In the low energy weak curvature regime, the nonminimal coupling
between the background geometry and the Higgs field can be neglected,
implying $\delta_{\rm cc}=1$.  For a
Friedmann-Lema\^{i}tre-Robertson-Walker (FLRW) space-time, the Weyl
tensor vanishes, hence the NCG corrections to the Einstein equation
vanish~\cite{Nelson:2008uy}, rending difficult to restrict $\beta$ (or
equivalently $\alpha_0$, or $f_0$) via cosmology or solar-system
tests. Imposing however a lower limit on $\beta$ is very important,
since it implies an upper limit to the moment $f_0$, corresponding to
a restriction on the particle physics at unification. This has been
achieved in Refs.~\cite{wjm1,wjm2}, by considering the energy lost to
gravitational radiation by orbiting binaries. 

Deriving the weak field limit of noncommutative spectral geometry, we
have shown~\cite{wjm1} that the production and dynamics of
gravitational waves are significantly altered and, in particular, the
graviton contains a massive mode that alters the energy lost to
gravitational radiation, in systems with evolving quadrupole moment.
Considering the rate of energy loss from a binary pair of masses $m_1,
m_2$, in the far field limit, we have shown~\cite{wjm1,wjm2} that the
orbital frequency
\be \omega = |\rho|^{-3/2} \sqrt{ G\left( m_1 + m_2\right)}~,  \ee
where $|\rho|$ stands for the magnitude of their separation vector,
has a critical value
\be
\label{critical}
2\omega_{\rm c} =\beta c~,
\ee
around which strong deviations from the familiar results of General
Relativity are expected. This maximum frequency results from the
natural length scale, given by $\beta^{-1}$, at which noncommutative
geometry effects become dominant.

The form of the gravitational radiation from binary systems can be
used to constrain $\beta$.  There are several binary pulsars for which
the rate of change of the orbital frequency has been well
characterised, and the predictions of General Relativity agree with
the data to high accuracy. Thus, on can restrict the parameter $\beta$
by requiring that the magnitude of deviations from General Relativity
be less than this uncertainty.  Requiring that $\beta > 2\omega /c$,
we have found~\cite{wjm2}
\be 
\label{constr-beta}
\beta > 7.55\times 10^{-13}~{\rm m}^{-1}~.
\ee
Since the strongest constraint comes from systems with high orbital
frequencies, one expects that future observations of rapidly orbiting
binaries, relatively close to the Earth, could improve it
by many orders of magnitude. 

\vskip.5truecm

Considering the background equations, the corrections to Einstein's
equations can only be apparent at leading order for anisotropic
models. Calculating the modified Friedmann equation for the Bianchi
type-V model, we have shown~\cite{Nelson:2008uy} that the correction
terms come in two types. The first one contains terms which are fourth
order in time derivatives, hence for the slowly varying functions
usually used in cosmology they can be neglected.  The second one
occurs at the same order as the standard Einstein-Hilbert terms,
however, it vanishes for homogeneous versions of Bianchi type-V. Thus,
although anisotropic cosmologies do contain corrections due to the
additional NCG terms in the action, they are typically of higher
order.  Inhomogeneous models do contain correction terms that appear
on the same footing as the ordinary (commutative) terms. In
conclusion, the corrections to Einstein's equations can only be
important for inhomogeneous and anisotropic space-times.

\vskip.5truecm

Certainly one cannot always neglect the coupling of the Higgs field to
the curvature. Namely, as energies approach the Higgs scale, this
nonminimal coupling can no longer be neglected, leading to corrections
even for background cosmologies. To understand the effects of these
corrections let us neglect the conformal term in Eq.~(\ref{eq:EoM2}),
so that the equations of motion read~\cite{Nelson:2008uy}
\be R^{\mu\nu} - \frac{1}{2}g^{\mu\nu}R =
\kappa_0^2\left[\frac{1}{1-\kappa_0^2 |{\bf H}|^2/6}\right] T^{\mu\nu}_{\rm
  matter}~. \ee 
Thus, $|{\bf H}|$ plays the r\^ole of an effective gravitational
constant~\cite{Nelson:2008uy}.

\vskip.5truecm 

The nonminimal coupling between the Higgs field and the Ricci
curvature may turn out to be particularly useful in early universe
cosmology~\cite{Nelson:2009wr,mmm}.  Such a coupling has been
introduced {\sl ad hoc} in the literature, in an attempt to drive
inflation through the Higgs field.  However, the value of the coupling
constant between the scalar field and the background geometry should
be dictated by the underlying theory.

In a FLRW metric, the Gravity-Higgs sector of the asymptotic
expansion of the spectral action, in Lorentzian
signature, reads
\be
S^{\rm
  L}_{\rm GH}=\int\Big[\frac{1-2\kappa_0^2\xi_0
    H^2}{2\kappa_0^2}R 
-\frac{1}{2}(\nabla  H)^2- V(H)\Big] \sqrt{-g}\  d^4x~,
\ee
where 
\be\label{higgs-pot}
V(H)=\lambda_0H^4-\mu_0^2H^2~,
\ee
with $\mu_0$ and $\lambda_0$ subject to radiative corrections as
functions of energy.  For large enough values of the Higgs field, the
renormalised value of these parameters must be calculated, while the
running of the top Yukawa coupling and the gauge couplings must be
evolved simultaneously.

At high energies the mass term is sub-dominant, and can be neglected.
For each value of the top quark mass, there is a value of the Higgs
mass where the effective potential is about to develop a metastable
minimum at large values of the Higgs field and the Higgs potential is
locally flattened~\cite{mmm}.  Since the region where the potential is
flat is narrow, the slow-roll must be very slow in order to provide a
sufficiently long period of quasi-exponential expansion.  Besides the
slow-roll parameters, denoted by $\epsilon$ and $\eta$, which may be
slow enough to allow sufficient number of e-folds, the amplitude of
density perturbations $\Delta_\mathcal{R}^2$ in the Cosmic Microwave
Background must be within the allowed experimental window. Inflation
predicts that at horizon crossing (denoted by stars), the amplitude of
density perturbations is related to the inflaton potential through
%q
\be \left(\frac{V_*}{\epsilon_*}\right)^{\frac14}
=2\sqrt{3\pi}\ m_\text{Pl}\ \Delta_\mathcal{R}^\frac12~,
\ee
where $\epsilon_*\leq1$.  Its value, as measured by
WMAP7~\cite{Larson:2010gs}, requires
\be \left(\frac{V_*}{\epsilon_*}\right)^{\frac14}
=(2.75\pm0.30)\times 10^{-2}\ m_\text{Pl}\,~,\label{eq:cobe} \ee
where $m_\text{Pl}$ stands for the Planck mass.  

Calculating~\cite{mmm} the renormalisation of the Higgs self-coupling
up to two-loops, we have constructed an effective potential which fits
the renormalisation group improved potential around the flat region.
We have found~\cite{mmm} a very good analytic fit to the Higgs
potential around the minimum of the potential:
\be
V^\text{eff}=\lambda_0^\text{eff}(H)H^4
=[a\ln^2(b\kappa H)+c] H^4~,
\ee
where the parameters $a, b$ are related to the low energy values of
top quark mass $m_{\rm t}$ as~\cite{mmm}
\beq
a(m_\text{t})&=&4.04704\times10^{-3}-4.41909\times10^{-5}
\left(\frac{m_\text{t}}{\text{GeV}}\right)
+1.24732\times10^{-7}\left(\frac{m_\text{t}}{\text{GeV}}\right)^2~,
\nonumber\\ 
b(m_\text{t})&=&\exp{\left[-0.979261
\left(\frac{m_\text{t}}{\text{GeV}}-172.051\right)\right]}~.
\eeq
The third parameter, $c$, encodes the appearance
of an extremum and depends on the values for top quark mass and Higgs
mass.  An extremum occurs if and only if $c/a\leq 1/16$, the
saturation of the bound corresponding to a perfectly flat region.  
It is convenient to write $c=[(1+\delta)/16]a$, where $\delta=0$
saturates the bound below which a local minimum is formed.  

This analysis was performed in the case of minimal coupling, so let us
investigate the modifications introduced in the case of a small
nonminimal coupling; within NCG the coupling is $\xi_0=1/12$.  We have
found~\cite{mmm} that the induced corrections to the potential imply
that flatness does not occur at $\delta=0$, but for fixed values of
$\delta$ depending on the value of the top quark mass. Thus, for
inflation to occur via the Higgs field, the top quark mass fixes the
Higgs mass extremely accurately.  Scanning carefully through the
parameter space, we concluded~\cite{mmm} that sufficient $e$-folds are
indeed generated provided a suitably tuned relationship between the
top quark mass and the Higgs mass holds. However, while the Higgs
potential can lead to the slow-roll conditions being satisfied once
the running of the self-coupling at two-loops is included, the
constraints imposed from the CMB data make the predictions of such a
scenario incompatible with the measured value of the top quark
mass~\cite{mmm}.  Running of the gravitational constant and
corrections by considering the more appropriate de\,Sitter, instead of
a Minkowski, background we found~\cite{mmm} that do not improve
substantially the realisation of a successful inflationary era.

%%%%%%%%%%%%%%%%
\section{Conclusions}
Noncommutative spectral geometry is a beautiful mathematical
construction with a rich phenomenological arena. In its present {\sl
  simple} form, which still remains classical and refers only to almost
commutative spaces, it offers an elegant explanation for the Standard
Model of electroweak and strong interactions.  The prediction for the
top quark mass is in agreement with current experimental data, while
the Higss mass is within the correct order of magnitude; its precise
value is excluded from the most recent experimental data but it is
still remarkable how close it remains to the experimental allowed
value besides the simplifications under which it was calculated.

Noncommutative spectral geometry lives by construction at very high
energy scales, offering a natural set-up to study early universe
cosmology. Late time astrophysics is a more difficult task due to
technical issues at the current stage of the NCG spectral
action~\cite{m2010}. Expecting further progress in computing exactly
the spectral action in its nonperturbative form and performing the
appropriate renormalisation group analysis, we expect that we will be
able to tackle astrophysical issues in the near future.

Here, I have reviewed possible cosmological fingerprints of
noncommutative spectral geometry, and proposed a mechanism to
constrain physics at unification through the implications of
production and dynamics of gravitational waves.

%%%%%%%%%%%%%%%%%%%%%%%%%%
\section{Acknowledgments}
It is a pleasure to thank the organisers of the 14th Conference on
Recent Developments in Gravity (NEB14)
``$N\epsilon\omega\tau\epsilon\rho\epsilon$s
E$\xi\epsilon\lambda\iota\xi\epsilon\iota$s $\sigma\tau\eta$
B$\alpha\rho\upsilon\tau\eta\tau\alpha$ 14'' in Ioannina
(Greece) for inviting me to present these results.  This work is
partially supported by the European Union through the Marie Curie
Research and Training Network {\sl UniverseNet} (MRTN-CT-2006-035863).

\end{document}